\documentclass[preprint,12pt]{elsarticle}

\usepackage{amsmath}
\usepackage{amssymb}
\usepackage{mathtools}
\usepackage{graphicx}
\usepackage{booktabs}
\usepackage{multirow}
\usepackage{tikz}
\usetikzlibrary{positioning,arrows.meta,decorations.pathreplacing,fit,calc}
\usepackage{url}
\usepackage[hidelinks]{hyperref}

\biboptions{sort&compress}
\makeatletter
\def\ps@pprintTitle{%
  \let\@oddhead\@empty
  \let\@evenhead\@empty
  \def\@oddfoot{\hfill\thepage\hfill}%
  \let\@evenfoot\@oddfoot}
\makeatother

\newcommand{\Sproj}{S_{\mathit{proj}}}
\newcommand{\Spm}{S_{\mathit{pm}}}
\newcommand{\Qneg}{Q_{\mathit{neg}}}
\newcommand{\QSproj}{Q_{S_{\mathit{proj}}}}
\newcommand{\team}{\mathit{team}}
\newcommand{\KV}{\mathit{KV}}
\newcommand{\rankeq}{\stackrel{\mathrm{rank}}{=}}
\newcommand{\fit}{\mathit{fit}}
\newcommand{\mng}{\mathit{mng}}
\newcommand{\com}{\mathit{com}}
\newcommand{\Mean}{\mathit{Mean}}

\begin{document}

\begin{frontmatter}

\title{Leader Finding: A New Variation of Expert Finding in Community Question Answering\tnoteref{t1}}
\tnotetext[t1]{A short preliminary version of this work appeared in the Proceedings of the 2021 ACM SIGIR International Conference on Theory of Information Retrieval (ICTIR '21), pages 281--285, \url{https://doi.org/10.1145/3471158.3472230}.}

\author[sbu]{Sepideh Entezari Maleki}
\ead{s.entezari@mail.sbu.ac.ir}

\author[sbu]{Mahmood Neshati\corref{cor1}}
\ead{m\_neshati@sbu.ac.ir}

\cortext[cor1]{Corresponding author}
\address[sbu]{Faculty of Computer Science and Engineering, Shahid Beheshti University, Tehran, Iran}

\begin{abstract}
Expert finding ranks people by their knowledge of a given topic. Many job positions need more than topical expertise, and a team leader is a clear case. A suitable leader knows the technical areas of the project without being overqualified, has strong managerial knowledge, and communicates well with the team. We study the leader finding problem, where the query is a project and a team and the output is a ranked list of candidate leaders. We model the problem in a probabilistic framework that combines technical fit, managerial knowledge, and communication with the team, and we propose one evaluation metric for each of these qualities. On two large collections built from Stack Overflow, the best configuration of our model more than doubles the NDCG@10 of the strongest baseline. These results show that a ranking of leaders must balance several qualities that topical expertise alone does not capture.
\end{abstract}


\end{frontmatter}

\section{Introduction}\label{sec:introduction}

Leaders play a major role in the success of a project. They guide the project in the right direction and lead the team in carrying out its tasks \citep{alvarenga2018revaluation,belassi1996new,verner2014factors}. Their roles include planning, defining tasks, motivating team members, and reducing conflict in the team \citep{alvarenga2018revaluation,belassi1996new,shastri2021role}. The skills that a leader needs fall into two groups, hard (technical) skills and soft (managerial) skills. Hard skills are the technical skill-areas in the scope of the project. A leader needs them for activities such as mentoring, assigning tasks, and controlling the quality of the project \citep{kerzner2017project,gillard2009soft}. Soft skills cover management and interpersonal communication, including motivating, monitoring, coordinating, and time management \citep{kerzner2017project,zaman2019understanding,gillard2009soft}. An effective leader balances the two groups to maximize team efficiency and reach the project goals \citep{alvarenga2018revaluation,kerzner2017project,gillard2009soft}.

Other factors, such as the team members, the organization \citep{belassi1996new}, time, and budget \citep{iriarte2020projects}, also affect the success of a project. Still, the choice of a suitable leader increases the chance of success \citep{iriarte2020projects,belassi1996new}. Expert finding methods can help find candidates for such a position. These methods estimate the depth of knowledge of each person in a topic from the documents associated with that person, and they rank people accordingly. However, most job positions, including a leadership position, depend on more than depth of knowledge. The level of knowledge, the cost of hiring, and the ability to communicate also matter.

Recent studies use the shape of expertise to match candidates to job positions \citep{gharebagh2018t,rostami2019t,dehghan2019temporal,dehghan2020mining}. Following \citet{gharebagh2018t}, these studies group people by their knowledge level (advanced or intermediate) across skill-areas. T-shaped people have advanced knowledge in one skill-area and intermediate knowledge in several skill-areas. Comb-shaped people have advanced knowledge in at least two skill-areas and intermediate knowledge in several skill-areas. Non-expert people have no advanced knowledge in any skill-area.

We study leader finding as a variation of expert finding that takes the shape of expertise into account. Our model ranks candidates by three conditions that a suitable leader should meet.
\begin{enumerate}
    \item \textbf{Technical fit.} A leader needs knowledge of the technical skill-areas in the scope of the project. Candidates with intermediate knowledge of these skill-areas and no knowledge outside them are the best fit. Leaders with a deep specialization in technical skills tend to become too involved in the technical side of the project, and this affects their management performance \citep{kerzner2017project}. Knowledge outside the project also points to overqualification, where a person has more skills than the job requires. Overqualification is associated with lower performance, lower commitment, frustration when rewards such as pay and promotion fall below expectations, and higher turnover intentions \citep{erdogan2011overqualified,harari2017thinks,erdogan2021overqualification}. Since organizations prefer affordable hires, overqualified candidates are a poor fit for the position.
    \item \textbf{Managerial knowledge.} Managerial skills such as motivating, time management, and task management are central to the role of a leader. The more managerial knowledge a candidate has, the more suitable the candidate is for the position.
    \item \textbf{Communication with the team.} A leader coaches and coordinates the team and helps solve its problems. These tasks require effective communication with the team members.
\end{enumerate}

\begin{figure}[t]
\centering
\begin{tikzpicture}[font=\small]
  \draw[thick, fill=gray!15] (0,2.2) rectangle (4.4,2.6);
  \foreach \x in {0.55,1.95,3.35} {
    \draw[thick, fill=teal!25] (\x,0) rectangle ++(0.5,2.2);
  }
  \draw[decorate, decoration={brace, amplitude=6pt}] (0,2.75) -- (4.4,2.75)
    node[midway, above=7pt, align=center] {Intermediate knowledge in the\\technical skill-areas of the project};
  \draw[decorate, decoration={brace, amplitude=6pt, mirror}] (0.55,-0.15) -- (3.85,-0.15)
    node[midway, below=7pt] {Advanced knowledge in managerial skill-areas};
  \begin{scope}[shift={(6.6,1.3)}]
    \draw[thick] (0,0) circle (0.8);
    \draw[thick] (0,0.25) circle (0.25);
    \draw[thick] (-0.45,-0.5) .. controls (-0.45,-0.05) and (0.45,-0.05) .. (0.45,-0.5);
    \node at (0,-1.15) {Comb-shaped leader};
  \end{scope}
\end{tikzpicture}
\caption{Shape of expertise of an ideal leader.}
\label{fig:c_shaped}
\end{figure}
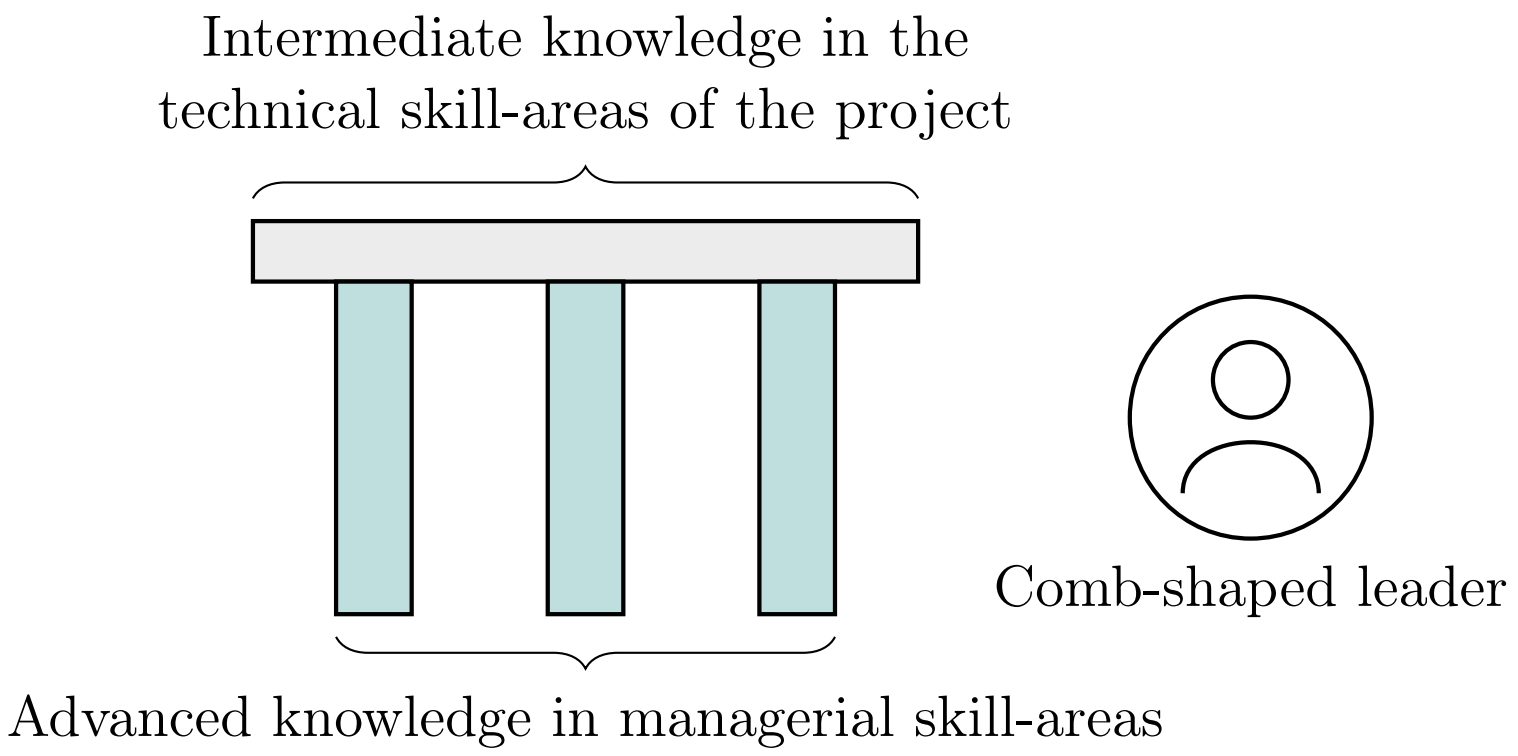

Under these conditions, the ideal candidate has intermediate knowledge in all required technical skill-areas of the project, advanced knowledge in managerial skill-areas, and no technical knowledge outside the project. By the definitions above, such a candidate is Comb-shaped, as Figure~\ref{fig:c_shaped} shows.

Existing expert finding methods do not model these conditions together. Our model combines the shape of expertise, technical and managerial knowledge, and graph-based features of communication. We evaluate it on collections built from Stack Overflow\footnote{\url{https://stackoverflow.com/}}, a large community question answering (CQA) site for software development, and we look for leaders of software development teams. Prior work has used the same data to find candidates for software teams \citep{gharebagh2018t,dehghan2020mining}.

In our preliminary work \citep{maleki2021novel}, we ranked candidates as the leader of a given team, where the team members vote for candidates based on centrality measures. In this article, we add the project to the query, model technical fit and managerial knowledge together with communication in one probabilistic framework, and report a broader set of evaluation metrics and experiments.

\subsection{Contributions}\label{sec:contributions}

Our main contributions are as follows.
\begin{enumerate}
    \item We formalize leader finding with a project and a team as the query. We model the problem as the estimation of a joint probability over technical fit, managerial knowledge, and communication with the team (Sections~\ref{sec:problem} and~\ref{sec:proposed}).
    \item We propose a negative knowledge model that adapts negative relevance feedback to penalize candidates with knowledge in technical skill-areas outside the project (Section~\ref{sec:nkm}).
    \item We propose a leader communication model that combines degree, closeness, and betweenness centrality in a communication graph of candidates and team members (Section~\ref{sec:lcm}).
    \item We propose three leader relevancy metrics for technical fit, managerial knowledge, and communication, together with their harmonic mean (Section~\ref{sec:leader_metrics}).
\end{enumerate}

\subsection{Organization}\label{sec:organization}

Section~\ref{sec:related} reviews related work. Section~\ref{sec:problem} defines the leader finding problem. Sections~\ref{sec:baselines} and~\ref{sec:proposed} present the baselines and the proposed model. Section~\ref{sec:experiments} describes the experimental setup and the results. Section~\ref{sec:discussion} discusses the findings and the limitations of our study, and Section~\ref{sec:conclusion} concludes the paper.

\section{Related Work}\label{sec:related}

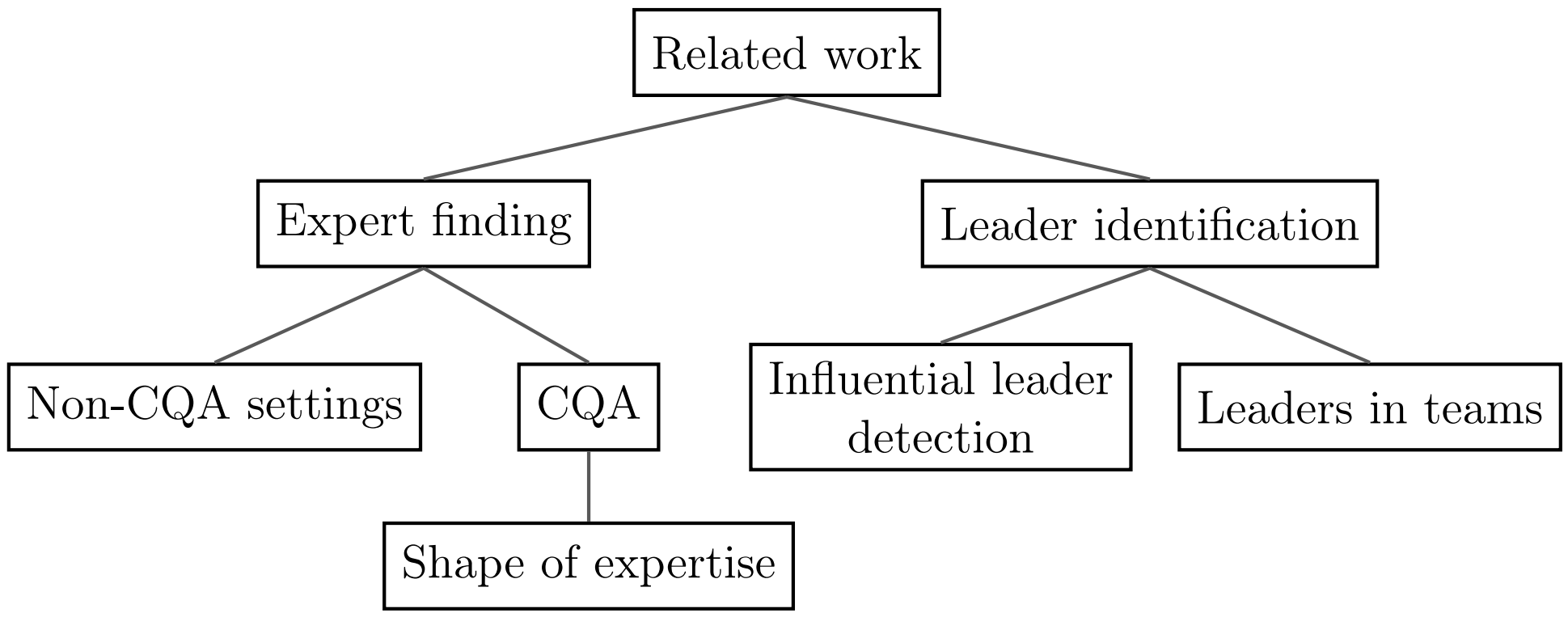
\begin{figure}[t]
\centering
\begin{tikzpicture}[font=\small, x=0.9cm,
    n/.style={draw, thick, align=center, inner sep=4pt, minimum height=7mm, fill=white},
    e/.style={thick, gray!70!black}]
  \node[n] (root) at (5.2,0) {Related work};
  \node[n] (ef) at (1.9,-1.4) {Expert finding};
  \node[n] (li) at (8.5,-1.4) {Leader identification};
  \node[n] (nc) at (0,-2.9) {Non-CQA settings};
  \node[n] (cqa) at (3.4,-2.9) {CQA};
  \node[n] (shape) at (3.4,-4.2) {Shape of expertise};
  \node[n] (inf) at (6.6,-2.9) {Influential leader\\detection};
  \node[n] (lit) at (10.5,-2.9) {Leaders in teams};
  \draw[e] (root.south) -- (ef.north);
  \draw[e] (root.south) -- (li.north);
  \draw[e] (ef.south) -- (nc.north);
  \draw[e] (ef.south) -- (cqa.north);
  \draw[e] (cqa.south) -- (shape.north);
  \draw[e] (li.south) -- (inf.north);
  \draw[e] (li.south) -- (lit.north);
\end{tikzpicture}
\caption{Categories of related work.}
\label{fig:related_work}
\end{figure}

We group related work into two areas, expert finding and leader identification (Figure~\ref{fig:related_work}).

\subsection{Expert Finding}\label{sec:related_ef}

Expert finding ranks candidates by their expertise in the topic of a query. It has been studied in academia \citep{cifariello2019wiser}, bibliographic networks \citep{neshati2012multi,daud2010temporal}, social networks \citep{lappas2009finding}, and CQA sites \citep{nobari2020quality,neshati2017dynamicity,kundu2021topic,zhou2014empirical,rostami2021intern}. Work on expert finding falls into CQA and non-CQA settings \citep{rostami2019t}. Since we find leaders on Stack Overflow, we focus on the CQA setting and on the shape of expertise.

Stack Overflow is one of the most popular CQA sites for software engineering, and many expert finding studies use CQA data. Some studies route questions to users who can answer them \citep{fu2020user,riahi2012finding,kundu2019formulation,li2019personalized}. Others find suitable people for a career opportunity \citep{gharebagh2018t,nobari2020quality,neshati2017dynamicity,rostami2021intern}. \citet{riahi2012finding} propose two statistical topic models for expert finding in CQA networks and compare them with traditional IR methods. \citet{neshati2017dynamicity} introduce the future expert finding problem and propose two learning to rank (LTR) models that predict who will become an expert in a topic. \citet{dargahi2017skill} propose two translation models that expand the query to reduce the vocabulary gap. The first model is probabilistic and uses mutual information, and the second learns an embedding space to translate each query.

These studies rank candidates by their expertise in the topics of the query. In contrast, we look for leaders who meet the conditions of a leadership position. Closer to our goal, several studies combine the level of knowledge with the shape of expertise to meet the needs of an organization. \citet{gharebagh2018t} introduce T-shaped expert finding in CQA sites, motivated by companies that follow agile software development. \citet{rostami2019t} form agile development teams for software projects. \citet{dehghan2019temporal} build temporal expertise profiles of users to detect T-shaped experts, and \citet{dehghan2020mining} propose a convolutional neural network that matches skill-areas and users through semantic vectors of their documents. \citet{rostami2021intern} rank Hyphen-shaped candidates for internship positions. Like these studies, we rank candidates for a given set of skills. However, a leader must also meet the managerial and communication conditions of Section~\ref{sec:introduction}, and our model covers these conditions as well.

\subsection{Leader Identification}\label{sec:related_leader}

Prior work on leaders either detects influential people in a network or deals with leaders within teams. Studies in the first group detect influential nodes for purposes such as spreading information in complex networks \citep{maji2020influential}, group recommendation \citep{nozari2020novel}, and online health communities \citep{zhao2014finding}. \citet{yang2020using} combine centrality measures with weighted k-means to identify leading firms in a venture capital network. These leaders influence the people or entities around them, but they are not selected for a given team or project.

Studies in the second group form teams that cover the requirements of a project and appoint one member as the leader, or they detect the leaders of existing teams. \citet{shi2017soc} form teams in social networks and decide during formation whether a new member joins as a leader or as a member. They use closeness centrality to keep the communication cost of the team low. \citet{dino2020detecting} detect leaders and key members of scientific teams in co-authorship networks.

Our work differs from the second group in three ways. First, these studies form teams or analyze existing teams, while we rank external candidates as the leader of a given team and project. Second, they select leaders mainly by their position in the network and do not consider technical knowledge, managerial knowledge, or fit to the required skills. We model these factors together with communication. Third, they do not define metrics that assess a selected leader against the requirements of a project and a team. We propose three such metrics.

\section{Problem Definition}\label{sec:problem}

In leader finding, the input is a project and a team. The project is a set of required technical skill-areas $\Sproj$, and the team is a set of members. The output is a ranked list of candidate leaders from a given pool of candidates. We prefer candidates who (1) cover all required technical skill-areas of the project with intermediate knowledge and have no technical knowledge outside the project, (2) have advanced knowledge in managerial skill-areas, and (3) communicate effectively with the team members. The first condition excludes overqualified candidates and candidates who would become too involved in the technical side of the project. Each candidate $e$ has a set of documents $D_e = D_{e,sa_1} \cup D_{e,sa_2} \cup \dots \cup D_{e,sa_m}$, where $D_{e,sa_i}$ is the set of documents that $e$ wrote in skill-area $sa_i$ and $m$ is the number of skill-areas in the collection. Section~\ref{sec:collection} explains how we extract skill-areas from our data.

\begin{figure}[t]
\centering
\includegraphics[width=0.85\textwidth]{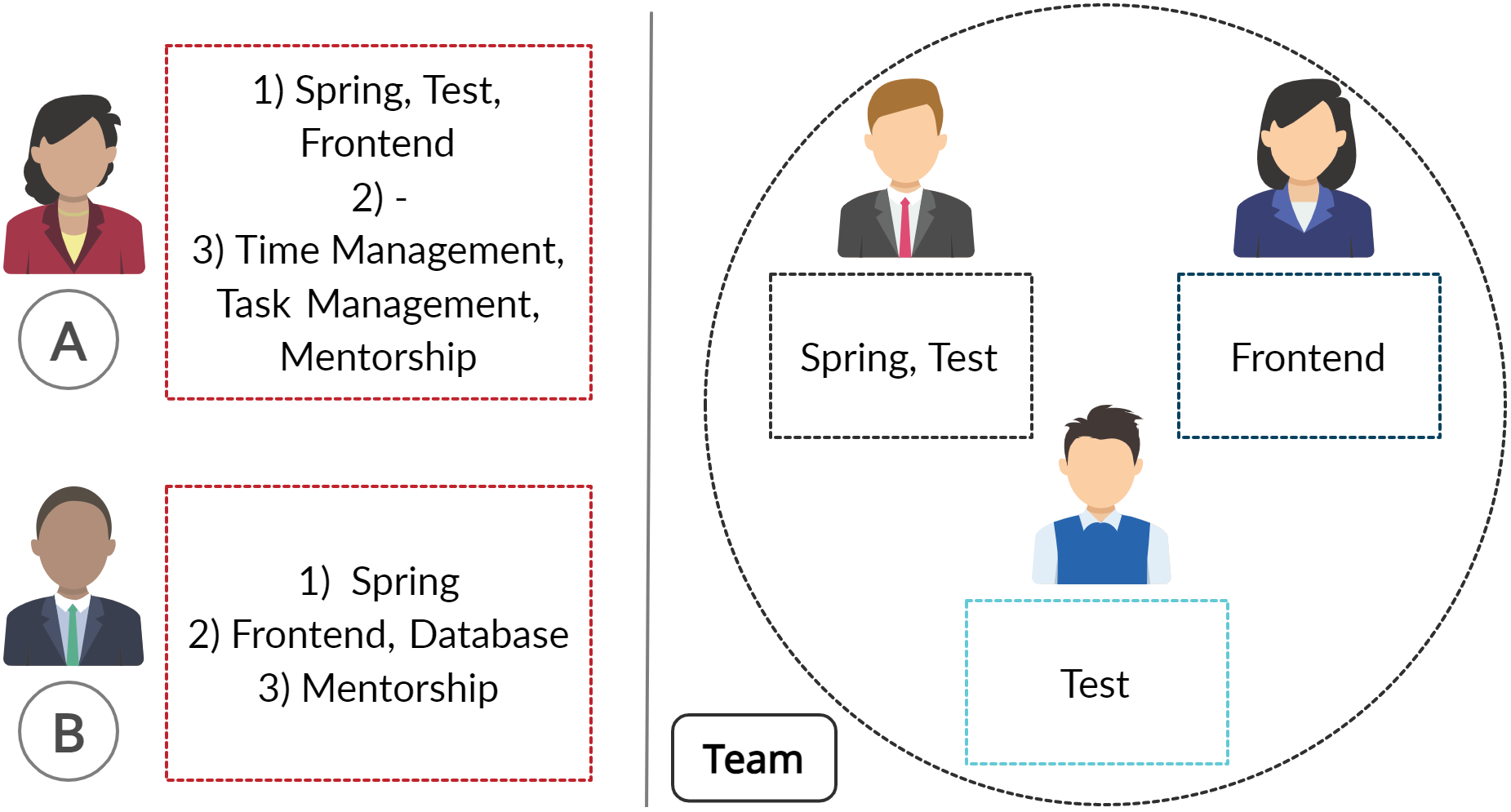}
\caption{Two candidate leaders, A and B, for a sample project and team. For each candidate, lines 1, 2, and 3 list the technical skill-areas with intermediate knowledge, the technical skill-areas with advanced knowledge, and the managerial skill-areas with advanced knowledge.}
\label{fig:problem_definition}
\end{figure}

Figure~\ref{fig:problem_definition} shows an example. The project requires \textit{Spring}, \textit{Frontend}, and \textit{Test}, and a team of three members covers these skill-areas. Candidate A is an ideal leader. She has intermediate knowledge in all three required skill-areas and no other technical knowledge, she shares skill-areas with every team member, and she has advanced knowledge in three managerial skill-areas. Candidate B should rank lower than A. He has intermediate knowledge only in \textit{Spring}, advanced knowledge in \textit{Frontend}, and knowledge in \textit{Database}, which the project does not require. He shares no skill-area with the member who works only on \textit{Test}, and he has less managerial knowledge than A.

We model leader finding with a probabilistic framework over the following events, each written as a random variable.
\begin{enumerate}
    \item $e$: candidate $e$ is selected as the leader.
    \item $\Sproj$: the selected candidate covers the required technical skill-areas $\Sproj$ with intermediate knowledge and has no knowledge outside them.
    \item $\Spm$: the selected candidate has advanced knowledge in the managerial skill-areas $\Spm$.
    \item $\team$: the selected candidate communicates effectively with the team members.
\end{enumerate}
The joint probability of these events,
\begin{equation}\label{eq:joint}
    P(e,\Sproj,\Spm,\team),
\end{equation}
measures how well $e$ meets all three conditions. Sections~\ref{sec:baselines} and~\ref{sec:proposed} present models that estimate it.

\section{Baseline Models}\label{sec:baselines}

Since leader finding has not been studied with this formulation before, we build three baselines from closely related expert finding models. Each baseline estimates Eq.~\eqref{eq:joint} under simplifying assumptions.

\subsection{Technical Knowledge Model (TKM)}\label{sec:tkm}

TKM ignores managerial knowledge and communication. It only looks for candidates with intermediate knowledge in the technical skill-areas of the project:
\begin{equation}\label{eq:tkm}
    P(e,\Sproj,\Spm,\team) \approx P(e,\Sproj),
\end{equation}
where $P(e,\Sproj)$ is the probability that $e$ has intermediate knowledge in $\Sproj$. We estimate it with the kernel-based model of \citet{rostami2019t}, which detects people with intermediate knowledge in given skill-areas. Under this model, a candidate is likely to have intermediate knowledge in skill-area $sa$ when the number of their documents in $sa$ is close to the average number of documents that candidates wrote in $sa$. We assume that knowledge in each skill-area of $\Sproj$ is independent of the other skill-areas:
\begin{equation}\label{eq:tkm_product}
    P(e,\Sproj) \approx \prod_{sa \in \Sproj} P(e,sa).
\end{equation}
To estimate $P(e,sa)$, we first compute the knowledge value of $e$ in $sa$:
\begin{equation}\label{eq:kv}
    \KV(e,sa) = \frac{\log\left(|D_{e,sa}| + 1\right)}{2 \log \Mean_{sa}},
\end{equation}
where $\Mean_{sa}$ is the average number of documents that candidates wrote in $sa$. A function $F_q$ then maps the knowledge value to $[0,1]$:
\begin{equation}\label{eq:fq}
    P(e,sa) = F_q\big(\KV(e,sa)\big) =
    \begin{cases}
        1 - \big(2\KV(e,sa) - 1\big)^2 & \text{if } \KV(e,sa) \leq 1,\\
        0 & \text{otherwise.}
    \end{cases}
\end{equation}
$F_q$ reaches its maximum of 1 when $|D_{e,sa}| + 1 = \Mean_{sa}$. It falls to 0 for candidates with no documents in $sa$ and for candidates with at least $\Mean_{sa}^2 - 1$ documents in $sa$, whose knowledge is beyond the intermediate level.

\subsection{Managerial Knowledge Model (MKM)}\label{sec:mkm}

MKM ignores technical knowledge and communication. It only looks for candidates with advanced knowledge in managerial skill-areas:
\begin{equation}\label{eq:mkm}
    P(e,\Sproj,\Spm,\team) \approx P(e,\Spm),
\end{equation}
where $P(e,\Spm)$ is the probability that $e$ has advanced knowledge in $\Spm$. We estimate it with the multi-expertise model of \citet{rostami2019t}, which detects people with advanced knowledge in given skill-areas. In this model, the more documents a candidate writes in a skill-area, the more likely the candidate is to have advanced knowledge in it. We assume that knowledge in each managerial skill-area is independent of the other skill-areas:
\begin{equation}\label{eq:mkm_product}
    P(e,\Spm) \approx \prod_{sa \in \Spm} P(e,sa) = \prod_{sa \in \Spm} P(sa)\,P(e \mid sa) \rankeq \prod_{sa \in \Spm} P(e \mid sa).
\end{equation}
The last step holds because $P(sa)$ does not depend on $e$. We estimate $P(e \mid sa)$ as
\begin{equation}\label{eq:mkm_smoothing}
    P(e \mid sa) = \lambda_d \frac{|D_{e,sa}|}{|D_{sa}|} + (1 - \lambda_d) \frac{|D_{sa}|}{|D|},
\end{equation}
where $D_{e,sa}$ is the set of documents that $e$ wrote in $sa$, $D_{sa}$ is the set of all documents in $sa$, $D$ is the set of all documents in the collection, and $\lambda_d \in [0,1]$ is a smoothing parameter. The second term keeps $P(e \mid sa)$ above zero for candidates with no documents in $sa$, so a single missing skill-area does not set the product in Eq.~\eqref{eq:mkm_product} to zero.

\subsection{Technical-Managerial Knowledge Model (TMKM)}\label{sec:tmkm}

TMKM ignores communication. It looks for candidates with advanced knowledge in managerial skill-areas and intermediate knowledge in the technical skill-areas of the project:
\begin{equation}\label{eq:tmkm}
    P(e,\Sproj,\Spm,\team) \approx P(e,\Sproj,\Spm) = P(\Sproj,\Spm \mid e)\,P(e).
\end{equation}
We assume that $\Sproj$ and $\Spm$ are conditionally independent given $e$:
\begin{equation}\label{eq:tmkm_independence}
    P(\Sproj,\Spm \mid e)\,P(e) = P(\Sproj \mid e)\,P(\Spm \mid e)\,P(e) = \frac{P(e,\Sproj)\,P(e,\Spm)}{P(e)},
\end{equation}
where $P(e)$ is the prior probability of selecting $e$. We assume that $P(e)$ is uniform, so it does not affect the ranking:
\begin{equation}\label{eq:tmkm_rank}
    P(e,\Sproj,\Spm) \rankeq P(e,\Sproj)\,P(e,\Spm).
\end{equation}
We estimate the two factors with TKM and MKM.

\section{Proposed Model}\label{sec:proposed}

The baselines miss two factors. They do not lower the rank of overqualified candidates with knowledge outside the project, and they do not model communication with the team. Our model adds both factors. We write Eq.~\eqref{eq:joint} as
\begin{equation}\label{eq:proposed}
    P(e,\Sproj,\Spm,\team) = P(\Sproj,\Spm,\team \mid e)\,P(e).
\end{equation}
As in TMKM, $P(e)$ is uniform. We further assume that $\Sproj$, $\Spm$, and $\team$ are conditionally independent given $e$:
\begin{equation}\label{eq:proposed_rank}
\begin{split}
    P(e,\Sproj,\Spm,\team) &\rankeq P(\Sproj \mid e)\,P(\Spm \mid e)\,P(\team \mid e)\\
    &\rankeq P(e,\Sproj)\,P(e,\Spm)\,P(e,\team).
\end{split}
\end{equation}
We estimate $P(e,\Spm)$ with MKM (Section~\ref{sec:mkm}), $P(e,\Sproj)$ with the negative knowledge model (NKM, Section~\ref{sec:nkm}), and $P(e,\team)$ with the leader communication model (LCM, Section~\ref{sec:lcm}).

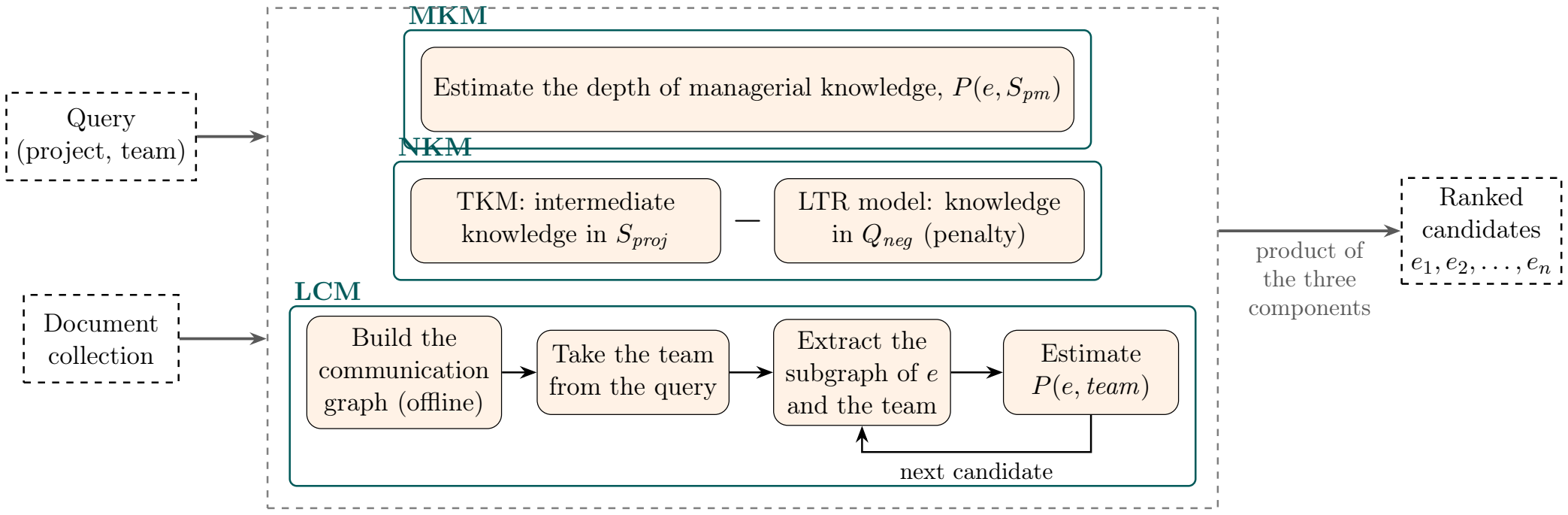
\begin{figure}[t]
\centering
\resizebox{\textwidth}{!}{%
\begin{tikzpicture}[font=\small, >={Stealth[length=2.5mm]},
    io/.style={draw, dashed, thick, align=center, minimum width=2.3cm, minimum height=1.3cm},
    st/.style={draw, rounded corners=6pt, align=center, fill=orange!10, minimum height=1.25cm, inner sep=5pt},
    comp/.style={draw, thick, teal!70!black, rounded corners=3pt, inner sep=7pt},
    lab/.style={font=\small\bfseries, teal!70!black, anchor=south west, inner sep=2pt}]
  \node[st, minimum width=7.6cm] (mkm) at (7.7,3.7) {Estimate the depth of managerial knowledge, $P(e,\Spm)$};
  \node[st, minimum width=4.6cm] (tkm) at (5.0,1.75) {TKM: intermediate\\knowledge in $\Sproj$};
  \node[st, minimum width=4.6cm] (neg) at (10.4,1.75) {LTR model: knowledge\\in $\Qneg$ (penalty)};
  \node at ($(tkm.east)!0.5!(neg.west)$) {\Large $-$};
  \node[st, minimum width=2.6cm] (g) at (2.6,-0.5) {Build the\\communication\\graph (offline)};
  \node[st, minimum width=2.6cm] (t) at (6.0,-0.5) {Take the team\\from the query};
  \node[st, minimum width=2.6cm] (s) at (9.4,-0.5) {Extract the\\subgraph of $e$\\and the team};
  \node[st, minimum width=2.6cm] (p) at (12.8,-0.5) {Estimate\\$P(e,\team)$};
  \draw[->, thick] (g) -- (t);
  \draw[->, thick] (t) -- (s);
  \draw[->, thick] (s) -- (p);
  \draw[->, thick] (p.south) -- ++(0,-0.55) -| node[pos=0.25, below, font=\footnotesize] {next candidate} (s.south);
  \node[comp, fit=(mkm)] (fm) {};
  \node[lab] at (fm.north west) {MKM};
  \node[comp, fit=(tkm)(neg)] (fn) {};
  \node[lab] at (fn.north west) {NKM};
  \node[comp, fit=(g)(p), inner ysep=14pt, yshift=-10pt] (fl) {};
  \node[lab] at (fl.north west) {LCM};
  \node[draw, dashed, thick, gray, inner sep=9pt, fit=(fm)(fn)(fl)] (model) {};
  \node[io] (query) at (-1.9,3.0) {Query\\(project, team)};
  \node[io] (docs) at (-1.9,0.0) {Document\\collection};
  \node[io] (out) at (18.6,1.6) {Ranked\\candidates\\$e_1, e_2, \dots, e_n$};
  \draw[->, very thick, gray!70!black] (query.east) -- (query.east -| model.west);
  \draw[->, very thick, gray!70!black] (docs.east) -- (docs.east -| model.west);
  \draw[->, very thick, gray!70!black] (out.west -| model.east) -- node[below, font=\footnotesize, align=center] {product of\\the three\\components} (out.west);
\end{tikzpicture}}
\caption{Architecture of the proposed model.}
\label{fig:arch}
\end{figure}

Figure~\ref{fig:arch} shows the architecture of the model. The input is a query, made of a project and a team, and the output is a ranked list of candidates. The three components use a document collection to estimate the knowledge of candidates. MKM estimates the managerial knowledge of each candidate. NKM estimates technical fit in two steps. It first estimates intermediate knowledge in the project skill-areas with TKM, and it then subtracts a penalty for knowledge outside the project that an LTR model estimates. LCM builds a communication graph once, independent of the query. For each query and each candidate, it extracts the subgraph that contains the candidate and the team and estimates $P(e,\team)$ from it. The final score of a candidate is the product of the three components.

\subsection{Negative Knowledge Model (NKM)}\label{sec:nkm}

NKM targets the first missing factor, technical fit. We treat knowledge in technical skill-areas outside the project as negative knowledge and penalize it. For this purpose, we adapt negative relevance feedback \citep{wang2008study}. In document retrieval, \citet{wang2008study} score a document $d$ for a query $Q$ with a negative query $\Qneg$ that contains terms that relevant documents should not contain:
$S_{\mathit{combined}}(Q,d) = S(Q,d) - \beta\, S(\Qneg,d)$, where $\beta$ controls the weight of the negative score. A document gets a high score when it matches $Q$ and does not match $\Qneg$. In the same way, we score candidate $e$ for the project as
\begin{equation}\label{eq:nkm}
    P(e,\Sproj) \propto P(e,\QSproj) - \beta\, P(e,\Qneg).
\end{equation}
$P(e,\QSproj)$ is the probability that $e$ has intermediate knowledge in $\Sproj$, and we estimate it with TKM (Section~\ref{sec:tkm}). $P(e,\Qneg)$ is the probability that $e$ has knowledge in the skill-areas of the negative query $\Qneg = S \setminus \Sproj$, where $S$ is the set of all technical skill-areas in the collection. This term penalizes knowledge outside the project. We apply min-max normalization over all candidates to map the score of Eq.~\eqref{eq:nkm} to $[0,1]$. We estimate $P(e,\Qneg)$ with LTR models, as described next.

\paragraph{Training data}
Each training example is a pair $(F_{e,\Qneg}, L)$, where $F_{e,\Qneg}$ is a feature vector for candidate $e$ and negative query $\Qneg$, and $L \in \{0,1\}$ indicates whether $e$ has knowledge in $\Qneg$. The feature vector $F_{e,\Qneg} = [f_{e,sa_1}, f_{e,sa_2}, \dots, f_{e,sa_n}]$ has one element for each technical skill-area $sa_i \in S$, where $n = |S|$:
\begin{equation}\label{eq:nkm_features}
    f_{e,sa_i} =
    \begin{cases}
        \dfrac{|D_{e,sa_i}| - \min_{sa_i}}{\max_{sa_i} - \min_{sa_i}} & \text{if } sa_i \in \Qneg,\\[2ex]
        0 & \text{otherwise,}
    \end{cases}
\end{equation}
where $\min_{sa_i}$ and $\max_{sa_i}$ are the minimum and maximum numbers of documents that candidates wrote in $sa_i$. The elements for the project skill-areas are zero, so the features only describe knowledge outside the project. We set $L=0$ if $e$ has neither intermediate nor advanced knowledge in any skill-area of $\Qneg$, and $L=1$ if $e$ has intermediate or advanced knowledge in at least one of them. A label of 1 therefore marks a candidate with knowledge outside the project.

\paragraph{Learning to rank models}
We estimate $P(e,\Qneg)$ with point-wise and pair-wise LTR methods. The point-wise methods treat the estimation as binary classification and score each candidate from its own feature vector, independent of the other candidates. We use a linear support vector machine (SVM-Linear), logistic regression (LR), and k-nearest neighbors (K-NN). The pair-wise methods compare two candidates at a time and learn to order each pair correctly. Their objective is to minimize the number of inversions, that is, pairs of candidates in the wrong order. We use SVM-Rank \citep{joachims2002optimizing} and RankNet \citep{burges2005learning}.

\subsection{Leader Communication Model (LCM)}\label{sec:lcm}

\begin{figure}[t]
\centering
\begin{tikzpicture}[font=\small,
    mem/.style={circle, fill=teal!70!black, text=white, minimum size=9mm},
    cand/.style={circle, fill=gray!55, minimum size=9mm},
    w/.style={fill=white, inner sep=1.5pt, font=\footnotesize}]
  \node[mem] (m1) at (0,2.6) {$M_1$};
  \node[mem] (m3) at (2.2,2.6) {$M_3$};
  \node[mem] (m2) at (4.4,2.6) {$M_2$};
  \node[cand] (c1) at (0.7,0) {$C_1$};
  \node[cand] (c2) at (3.7,0) {$C_2$};
  \draw[thick, gray!70!black] (m1) -- node[w] {0.5} (m3);
  \draw[thick, gray!70!black] (c1) -- node[w] {0} (m1);
  \draw[thick, gray!70!black] (c1) -- node[w, pos=0.5] {0} (m3);
  \draw[thick, gray!70!black] (c1) -- node[w, pos=0.22] {0} (m2);
  \draw[thick, gray!70!black] (c2) -- node[w, pos=0.22] {0.5} (m1);
  \draw[thick, gray!70!black] (c2) -- node[w] {0} (m2);
\end{tikzpicture}
\caption{A sample communication graph with team members ($M_1$, $M_2$, $M_3$) and candidate leaders ($C_1$, $C_2$). Edge weights are communication costs.}
\label{fig:graph}
\end{figure}
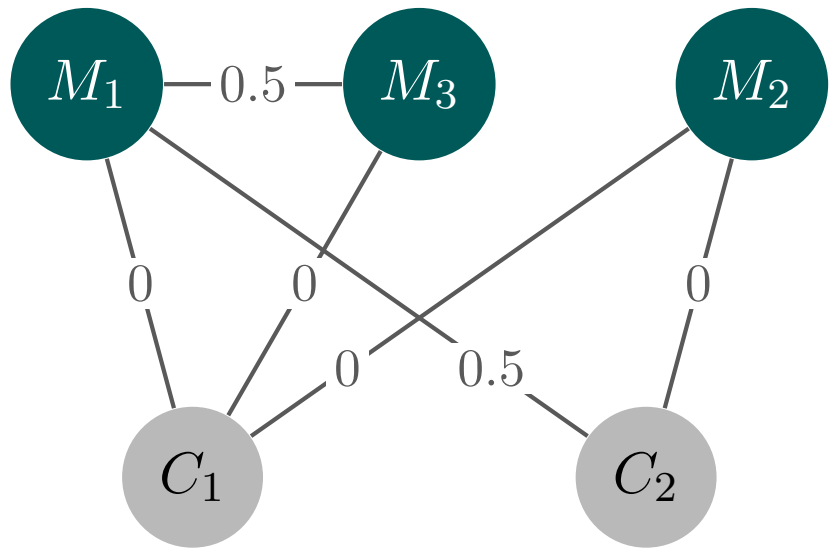

LCM targets the second missing factor, communication between a candidate and the team, and estimates $P(e,\team)$. We infer the strength of the association between a candidate and the team members from a communication graph $G=(V,E)$. The vertices are users, who act as candidates and team members, and the edges are their relations. $G$ is undirected and weighted, and we build it once, independent of the query. Team formation studies use similar graphs \citep{lappas2009finding}. The weight of an edge is the communication cost between its two users, and a lower weight means easier communication. The distance between two users is the minimum total weight of a path between them. Section~\ref{sec:graph} describes how we build $G$ from Stack Overflow. Figure~\ref{fig:graph} shows an example with three team members, $M_1$, $M_2$, and $M_3$, and two candidates, $C_1$ and $C_2$.

We estimate $P(e,\team)$ with the centrality measures of \citet{freeman1978centrality}. Several studies report that a leader in a central position in the team improves coordination and communication \citep{ehrlich2014communication,nootjarat2015moderating,chiu2017managers,alves2020does}. A central leader can share information, mentor team members, and help solve problems in the team. We combine degree, closeness, and betweenness centrality:
\begin{equation}\label{eq:lcm}
    P(e,\team) = \lambda_{dc} f_{dc}(e) + \lambda_{cc} f_{cc}(e) + \lambda_{bc} f_{bc}(e),
\end{equation}
where each $f_{\bullet}(e)$ is a centrality score and the weights satisfy $\lambda_{dc} + \lambda_{cc} + \lambda_{bc} = 1$. Let $G_{e,\team} = (V',E')$ be the subgraph of $G$ that contains candidate $e$, the team members, and the edges among them.

Degree centrality $f_{dc}(e)$ measures how strongly $e$ is connected to the team members. Intuitively, the best candidate has the strongest direct links to the team:
\begin{equation}\label{eq:degree}
    f_{dc}(e) = \frac{\sum_{v \in V' \setminus \{e\}} w_{e,v}}{|V'| - 1},
\end{equation}
where $w_{e,v} = c_{\max} - c_{e,v}$ if $e$ and $v$ are adjacent and $w_{e,v} = 0$ otherwise. Here, $c_{e,v}$ is the communication cost of the edge between $e$ and $v$, and $c_{\max}$ is the maximum communication cost in $G$.

Closeness centrality $f_{cc}(e)$ measures how fast $e$ reaches the team members, so candidates at a small distance from the team rank higher:
\begin{equation}\label{eq:closeness}
    f_{cc}(e) = \frac{1}{\sum_{v \in V' \setminus \{e\}} d_{e,v}},
\end{equation}
where $d_{e,v}$ is the shortest-path distance between $e$ and $v$ in $G$.

Betweenness centrality $f_{bc}(e)$ measures how often $e$ lies on shortest paths that start at a team member. A candidate on many such paths is easy to reach from the team:
\begin{equation}\label{eq:betweenness}
    f_{bc}(e) = \sum_{v \in V' \setminus \{e\}} \; \sum_{u \in V \setminus \{e, v\}} \frac{g_{v,u}(e)}{g_{v,u}},
\end{equation}
where $g_{v,u}$ is the number of shortest paths between $v$ and $u$ in $G$ and $g_{v,u}(e)$ is the number of those paths that pass through $e$.

We learn the weights $\lambda_{\bullet}$ with coordinate ascent (CA) \citep{metzler2007linear}, which optimizes one parameter at a time while keeping the others fixed. Each training example is the vector of the three centrality scores of a pair $(e,\team)$. We label it 1 if the communication score $\com(e)$ of Section~\ref{sec:leader_metrics} is above 0.5 and 0 otherwise. The learned weights show how much each centrality measure contributes to communication with the team. We learn them with 5-fold cross-validation over queries, so the weights applied to a query are never learned from that query (Section~\ref{sec:queries}).

\section{Experimental Setup and Results}\label{sec:experiments}

This section describes the test collections (Section~\ref{sec:collection}), the communication graph (Section~\ref{sec:graph}), the evaluation metrics (Section~\ref{sec:metrics}), the parameter settings (Section~\ref{sec:parameters}), and the queries and implementation (Section~\ref{sec:queries}). Section~\ref{sec:results} reports the results.

\subsection{Test Collections}\label{sec:collection}

We find leaders for software development projects and teams. Our data contains all Stack Overflow posts from August 2008 to March 2015, which prior work uses for expert finding in CQA sites \citep{gharebagh2018t,dehghan2020mining}. We build two technical collections from all questions tagged \textit{Java} or \textit{C\#} and their answers, and one managerial collection from all posts tagged \textit{project-management}. Table~\ref{tab:statistics} reports their statistics.

\begin{table}[t]
\centering
\caption{Statistics of the test collections.}
\label{tab:statistics}
\begin{tabular}{lrrr}
\toprule
\textbf{Collection} & \textbf{\#Questions} & \textbf{\#Answers} & \textbf{\#Users} \\
\midrule
Java & 810,071 & 1,512,812 & 83,557 \\
C\# & 763,717 & 1,453,649 & 84,095 \\
Project management & 1,746 & 7,083 & 3,642 \\
\bottomrule
\end{tabular}
\end{table}

Our evaluation needs a ground truth that gives the knowledge level of each user in each skill-area. We adapt the pipeline of \citet{gharebagh2018t} to build it, as described below.

\subsubsection{Extracting Skill-Areas}\label{sec:skill_areas}

\begin{figure}[t]
\centering
\begin{tabular}{cc}
\includegraphics[width=0.44\textwidth]{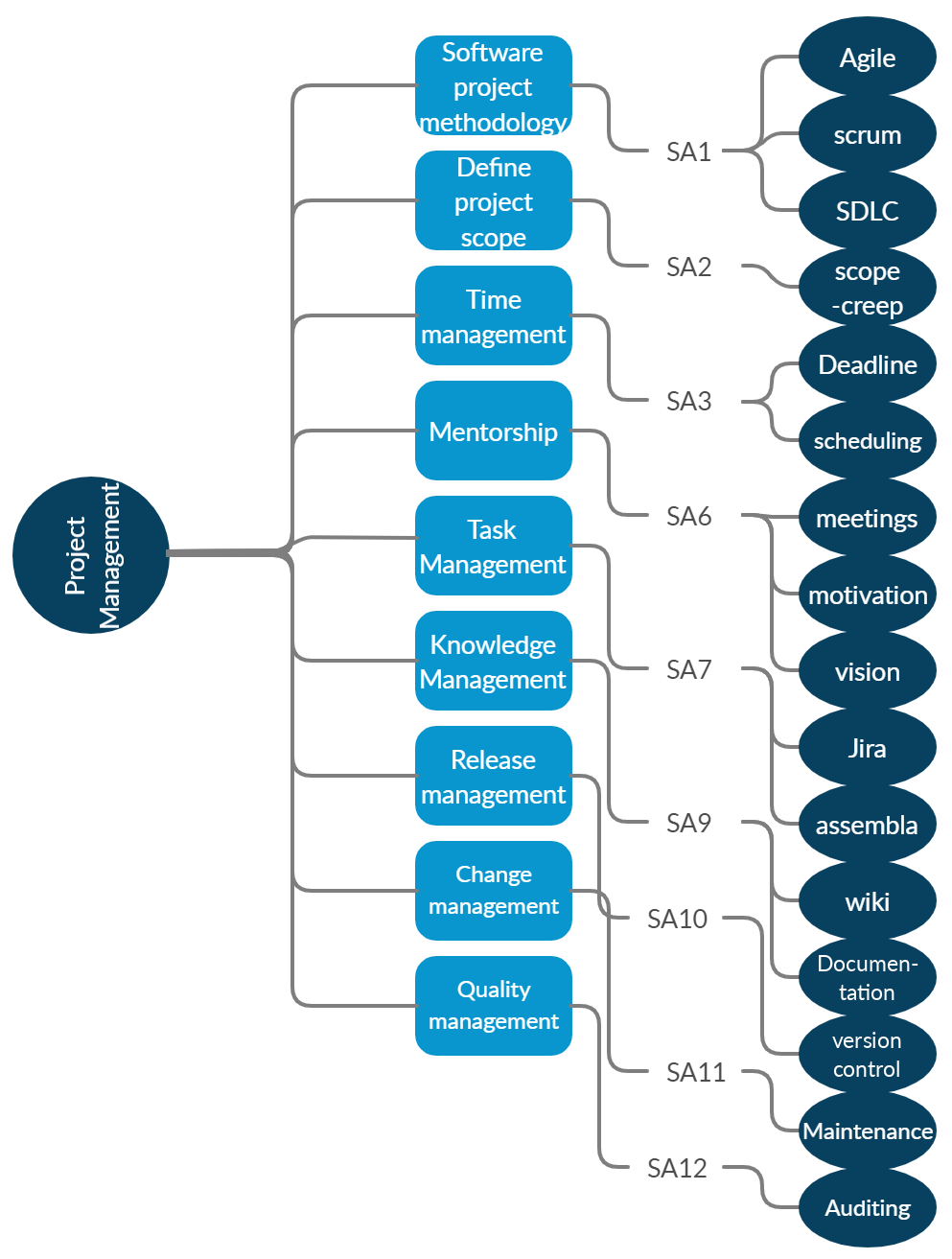} &
\includegraphics[width=0.44\textwidth]{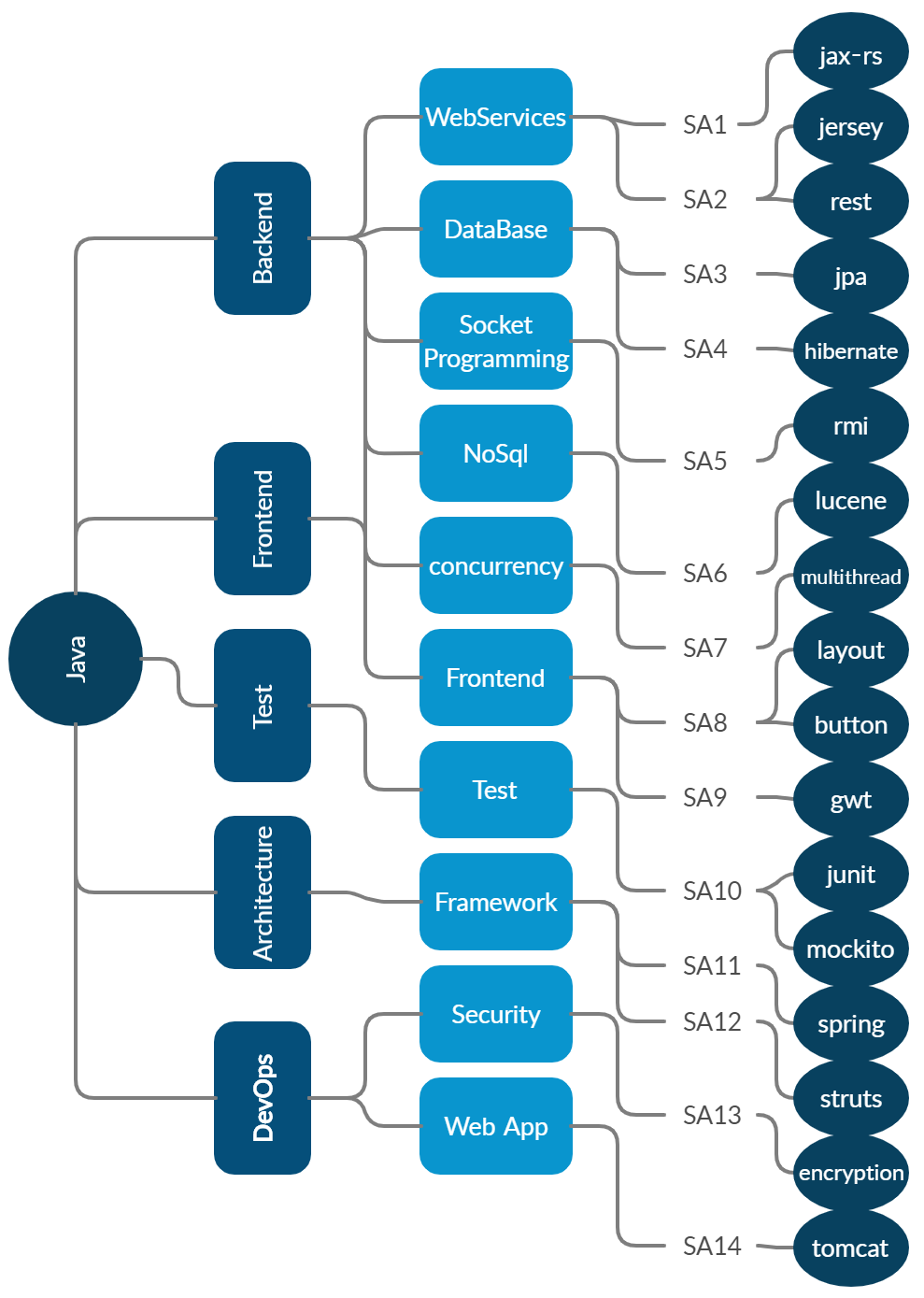} \\
(a) Managerial domain & (b) Java domain
\end{tabular}
\caption{Part of the skill-area trees of (a) the managerial domain and (b) the Java domain. From left to right, the levels are domain, team position (Java only), skill-area, cluster label (SA$i$), and tag.}
\label{fig:trees}
\end{figure}

Each Stack Overflow question has one or more tags, and we derive skill-areas from these tags. Unlike \citet{gharebagh2018t}, we organize skill-areas in a multi-level tree. Figure~\ref{fig:trees} shows part of the tree for the managerial domain and for the Java domain. The technical tree has five levels, from the leaves to the root.
\begin{enumerate}
    \setcounter{enumi}{4}
    \renewcommand{\labelenumi}{(\arabic{enumi})}
    \item \textbf{Tags.} The most frequent tags of the collection.
    \addtocounter{enumi}{-2}
    \item \textbf{Cluster labels.} We group tags with agglomerative hierarchical clustering with average linkage, so tags that often occur in the same post fall into the same cluster. Each node at this level is the label of one cluster.
    \addtocounter{enumi}{-2}
    \item \textbf{Skill-areas.} We refine the clusters of level 4 into skill-areas. The Java and C\# collections have 27 and 21 skill-areas.
    \addtocounter{enumi}{-2}
    \item \textbf{Team positions.} The positions of a software development team, which two recruiters assign to the skill-areas. The collections have five team positions.
    \addtocounter{enumi}{-2}
    \item \textbf{Domain.} The name of the domain.
\end{enumerate}
To check the reliability of the assignment to team positions, we measure the inter-rater agreement \citep{campbell2013coding} between the two recruiters with Cohen's $\kappa$. The recruiters agree on 24 of the 27 Java skill-areas ($\kappa = 0.85$) and on 19 of the 21 C\# skill-areas ($\kappa = 0.87$). Both values indicate strong agreement \citep{mchugh2012interrater}. The managerial tree has the same structure without the team-position level, because a leader, not a team, needs to cover the managerial skill-areas.

\subsubsection{Knowledge Levels}\label{sec:knowledge_levels}

Following \citet{gharebagh2018t}, we determine the knowledge level of a user in a skill-area from the number of their accepted answers in that skill-area, and we group users into beginner, intermediate, and advanced levels. From these levels across skill-areas, we label each user as non-expert, T-shaped, or Comb-shaped (Section~\ref{sec:introduction}).

\subsection{Communication Graph}\label{sec:graph}

We build one communication graph $G=(V,E)$ for each technical domain, Java and C\#. $V$ contains all users of the collection. We connect two users $u$ and $v$ when they have documents in at least one common skill-area, and we set the weight of the edge to $1 - \mathit{Jaccard}(u,v)$, which is the communication cost of Section~\ref{sec:lcm}. We use the weighted Jaccard similarity \citep{ioffe2010improved}:
\begin{equation}\label{eq:jaccard}
    \mathit{Jaccard}(u,v) = \frac{\sum_{i=1}^{m} \min\left(n_u^i, n_v^i\right)}{\sum_{i=1}^{m} \max\left(n_u^i, n_v^i\right)},
\end{equation}
where $n_u^i$ and $n_v^i$ are the numbers of documents that $u$ and $v$ wrote in the $i$-th skill-area.

\subsection{Evaluation Metrics}\label{sec:metrics}

We first define three leader relevancy metrics that score a single candidate. We then use them as grades in standard ranking metrics.

\subsubsection{Leader Relevancy Metrics}\label{sec:leader_metrics}

A suitable leader (1) fits the project in technical skill-areas, (2) has advanced managerial knowledge, and (3) communicates well with the team. We define one metric for each quality, called fittingness, managerial, and communication, and we combine them with their harmonic mean. For a user $e$, let $T_e$ be the set of technical skill-areas and $M_e$ the set of managerial skill-areas in which $e$ has intermediate or advanced knowledge.

\paragraph{Fittingness}
Fittingness measures how well a candidate covers the technical skill-areas of the project. A fit candidate has intermediate knowledge in exactly these skill-areas. Knowledge in other technical skill-areas, at either level, lowers the score. We build fittingness on graded precision (GP) and graded recall (GR) \citep{kekalainen2002using}. In document retrieval, for a set of retrieved documents $D$ and a set of relevant documents $R$,
\begin{equation}\label{eq:gp_gr}
    GP(D) = \frac{\sum_{d \in D} rel_d}{|D|}, \qquad GR(D) = \frac{\sum_{d \in D} rel_d}{\sum_{d \in R} rel_d},
\end{equation}
where $rel_d \in [0,1]$ is the relevance grade of document $d$. For leader finding, GR measures how well the candidate covers the project skill-areas with intermediate knowledge, and GP penalizes knowledge outside the project:
\begin{equation}\label{eq:fit_gp_gr}
    GP(T_e) = \frac{\sum_{sa \in T_e} rel_{sa}}{|T_e|}, \qquad GR(T_e) = \frac{\sum_{sa \in T_e} rel_{sa}}{|\Sproj|},
\end{equation}
\begin{equation}\label{eq:fit_rel}
    rel_{sa} =
    \begin{cases}
        1 & \text{if } sa \in \Sproj \text{ and } e \text{ has intermediate knowledge in } sa,\\
        0.5 & \text{if } sa \in \Sproj \text{ and } e \text{ has advanced knowledge in } sa,\\
        0 & \text{if } sa \notin \Sproj.
    \end{cases}
\end{equation}
Advanced knowledge in a project skill-area gets half credit, because the position needs intermediate rather than deep technical knowledge. Fittingness is the harmonic mean of GP and GR:
\begin{equation}\label{eq:fit}
    \fit(e) = \frac{2 \cdot GP(T_e) \cdot GR(T_e)}{GP(T_e) + GR(T_e)},
\end{equation}
and $\fit(e) = 0$ when $T_e$ is empty or $GP(T_e) + GR(T_e) = 0$.

\paragraph{Managerial}
The managerial metric measures how well a candidate covers the managerial skill-areas, with a preference for advanced knowledge. We again use GP and GR:
\begin{equation}\label{eq:mng_gp_gr}
    GP(M_e) = \frac{\sum_{sa \in M_e} rel_{sa}}{|M_e|}, \qquad GR(M_e) = \frac{\sum_{sa \in M_e} rel_{sa}}{|\Spm|},
\end{equation}
where $\Spm$ is the set of all managerial skill-areas and $rel_{sa}$ is 1 for advanced and 0.5 for intermediate knowledge. The managerial metric $\mng(e)$ is the harmonic mean of $GP(M_e)$ and $GR(M_e)$, defined as in Eq.~\eqref{eq:fit}.

\paragraph{Communication}
The communication metric measures how well a candidate can communicate with the team. A leader who knows the skill-areas of the team members can discuss their work with them and coach and coordinate them. We measure the overlap between the technical skill-areas of $e$ and those of each member $u$:
\begin{equation}\label{eq:pcom}
    P_{\com}(e,u) = \frac{|T_e \cap T_u|}{|T_u|}, \qquad \com(e) = \frac{\sum_{u \in \team} P_{\com}(e,u)}{|\team|}.
\end{equation}

\paragraph{F-measure}
The three metrics trade off against each other. To compare candidates on all three, we use their harmonic mean:
\begin{equation}\label{eq:fmeasure}
    F(e) = \frac{3}{\frac{1}{\fit(e)} + \frac{1}{\mng(e)} + \frac{1}{\com(e)}},
\end{equation}
with $F(e) = 0$ when any of the three scores is zero.

\begin{figure}[t]
\centering
\includegraphics[width=\textwidth]{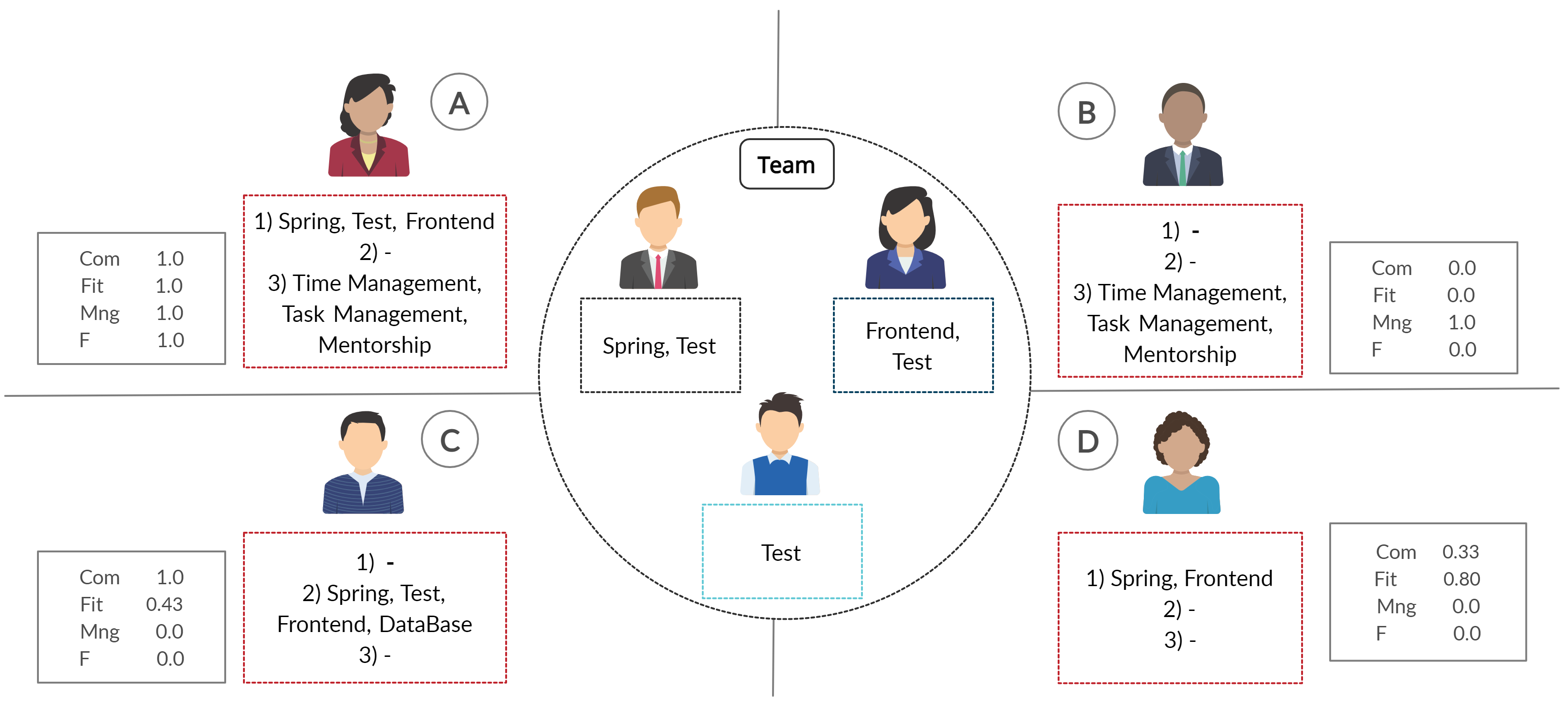}
\caption{Leader relevancy metrics for four sample candidates. For each candidate, lines 1, 2, and 3 list the technical skill-areas with intermediate knowledge, the technical skill-areas with advanced knowledge, and the managerial skill-areas with advanced knowledge. Com, Fit, Mng, and F are the communication, fittingness, managerial, and F-measure scores.}
\label{fig:metrics_example}
\end{figure}

Figure~\ref{fig:metrics_example} illustrates the metrics for a project with \textit{Spring}, \textit{Frontend}, and \textit{Test} and a team of three members who all know \textit{Test}. Candidate A is ideal. She has intermediate knowledge in all three project skill-areas and no other technical knowledge, advanced knowledge in all managerial skill-areas of the example, and shared skill-areas with every member. All her scores are 1. Candidate B has advanced managerial knowledge but no technical knowledge, so his managerial score is 1 and his fittingness and communication scores are 0. Candidate C has advanced knowledge in all three project skill-areas, which gives him a communication score of 1. However, his technical knowledge is deeper than the position needs, and he also knows \textit{Database}, which the project does not require. His fittingness is therefore 0.43. Candidate D has intermediate knowledge in \textit{Spring} and \textit{Frontend} and no knowledge outside the project, so her fittingness is 0.80. She does not know \textit{Test}, which all three members know, and her communication score is 0.33. Since C and D have no managerial knowledge, their F-measure is 0.

\subsubsection{Ranking Metrics}\label{sec:ranking_metrics}

To evaluate a ranked list of candidates, we use normalized discounted cumulative gain (NDCG) \citep{jarvelin2002cumulated} and graded average precision (GAP) \citep{robertson2010extending}. Both metrics accept graded relevance, and we use the leader relevancy metrics of Section~\ref{sec:leader_metrics} as grades. For example, to evaluate a ranking by fittingness, we use the fittingness score of each candidate as its grade. Unless we state otherwise, NDCG@$k$ and GAP@$k$ use the F-measure as the grade.

\subsection{Parameter Settings}\label{sec:parameters}

We set $\lambda_d = 0.9$ in MKM and TMKM, following \citet{rostami2019t}. For NKM, we tried several values of $\beta$ and set $\beta = 0.2$. We learn the weights of LCM with CA (Section~\ref{sec:lcm}). The learned weights are $\lambda_{dc} = 0.60$, $\lambda_{cc} = 0.23$, and $\lambda_{bc} = 0.17$ for Java, and $\lambda_{dc} = 0.57$, $\lambda_{cc} = 0.24$, and $\lambda_{bc} = 0.19$ for C\#.

\subsection{Queries and Implementation}\label{sec:queries}

A query consists of a project and a team. A project is a set of $k$ technical skill-areas that covers every team position with at least one skill-area. Since the collections have five team positions, $k \geq 5$, and we use $k = 5$ in all experiments. Each team member covers one skill-area of the project, so a team has five members. To build teams for a project, we randomly select three users with knowledge in each of its skill-areas. We then form teams by choosing one of the three users for each skill-area, which gives up to $3^k = 243$ teams for each project. Each pair of a project and a team is one query. This process gives 6,804 queries for Java and 1,512 queries for C\#. For each query, all users of the collection except the team members are candidates.

We use 5-fold cross-validation over the queries to train the learning-based NKM variants and to learn the LCM weights with CA. We use RankLib\footnote{\url{https://sourceforge.net/p/lemur/wiki/RankLib/}} for RankNet and coordinate ascent, the SVM-Rank implementation of Joachims\footnote{\url{https://www.cs.cornell.edu/people/tj/svm_light/svm_rank.html}} with a linear kernel and $C = 20$, and scikit-learn\footnote{\url{https://scikit-learn.org/}} for the point-wise models.

\subsection{Results}\label{sec:results}

We answer three research questions.
\begin{itemize}
    \item \textbf{RQ1:} How do the baselines compare on the three leader relevancy metrics, and how well does each baseline balance them?
    \item \textbf{RQ2:} How do the variants of the negative knowledge model compare?
    \item \textbf{RQ3:} How does the proposed model compare with the baselines?
\end{itemize}
Sections~\ref{sec:rq1} to~\ref{sec:rq3} answer them in turn.

\begin{table}[t]
\centering
\caption{Comparison of the proposed model with the baselines, with the F-measure as the grade. $^{*}$ marks a baseline that significantly outperforms the other baselines, and $^{**}$ marks a significant improvement over all baselines ($p < 0.05$). The best value in each row is in bold.}
\label{tab:main}
\small
\setlength{\tabcolsep}{4pt}
\begin{tabular}{llccccc}
\toprule
 & & \multicolumn{3}{c}{\textbf{Baselines}} & \multicolumn{2}{c}{\textbf{Proposed}} \\
\cmidrule(lr){3-5}\cmidrule(lr){6-7}
\textbf{Collection} & \textbf{Metric} & \textbf{TKM} & \textbf{MKM} & \textbf{TMKM} & \textbf{Point-wise} & \textbf{Pair-wise} \\
\midrule
\multirow{4}{*}{C\#}
 & NDCG@10 & 0.048 & 0.165 & 0.236$^{*}$ & \textbf{0.621}$^{**}$ & 0.534$^{**}$ \\
 & NDCG@20 & 0.059 & 0.203 & 0.253$^{*}$ & \textbf{0.650}$^{**}$ & 0.598$^{**}$ \\
 & GAP@10  & 0.018 & 0.071 & 0.112$^{*}$ & \textbf{0.465}$^{**}$ & 0.424$^{**}$ \\
 & GAP@20  & 0.016 & 0.106 & 0.131$^{*}$ & \textbf{0.476}$^{**}$ & 0.433$^{**}$ \\
\midrule
\multirow{4}{*}{Java}
 & NDCG@10 & 0.005 & 0.216 & 0.280$^{*}$ & \textbf{0.572}$^{**}$ & 0.544$^{**}$ \\
 & NDCG@20 & 0.006 & 0.262 & 0.314$^{*}$ & \textbf{0.568}$^{**}$ & 0.558$^{**}$ \\
 & GAP@10  & 0.001 & 0.144 & 0.178$^{*}$ & \textbf{0.452}$^{**}$ & 0.385$^{**}$ \\
 & GAP@20  & 0.000 & 0.149 & 0.183$^{*}$ & \textbf{0.483}$^{**}$ & 0.392$^{**}$ \\
\bottomrule
\end{tabular}
\end{table}

\begin{table}[t]
\centering
\caption{Comparison of the negative knowledge model variants, with fittingness as the grade. The best value in each row is in bold.}
\label{tab:nkm}
\small
\setlength{\tabcolsep}{3.5pt}
\resizebox{\textwidth}{!}{%
\begin{tabular}{llccccc}
\toprule
 & & \multicolumn{3}{c}{\textbf{Point-wise}} & \multicolumn{2}{c}{\textbf{Pair-wise}} \\
\cmidrule(lr){3-5}\cmidrule(lr){6-7}
\textbf{Collection} & \textbf{Metric} & \textbf{SVM-Linear} & \textbf{LR} & \textbf{K-NN} & \textbf{SVM-Rank} & \textbf{RankNet} \\
\midrule
\multirow{4}{*}{C\#}
 & NDCG@10 & 0.301 & 0.322 & \textbf{0.368} & 0.282 & 0.264 \\
 & NDCG@20 & 0.288 & 0.312 & \textbf{0.350} & 0.265 & 0.231 \\
 & GAP@10  & 0.231 & 0.258 & \textbf{0.346} & 0.204 & 0.168 \\
 & GAP@20  & 0.195 & 0.223 & \textbf{0.291} & 0.161 & 0.121 \\
\midrule
\multirow{4}{*}{Java}
 & NDCG@10 & 0.496 & 0.496 & \textbf{0.529} & 0.486 & 0.463 \\
 & NDCG@20 & 0.468 & 0.469 & \textbf{0.504} & 0.461 & 0.437 \\
 & GAP@10  & 0.381 & 0.381 & \textbf{0.448} & 0.367 & 0.348 \\
 & GAP@20  & 0.316 & 0.317 & \textbf{0.361} & 0.307 & 0.290 \\
\bottomrule
\end{tabular}}
\end{table}

\begin{figure}[t]
\centering
\includegraphics[width=0.49\textwidth]{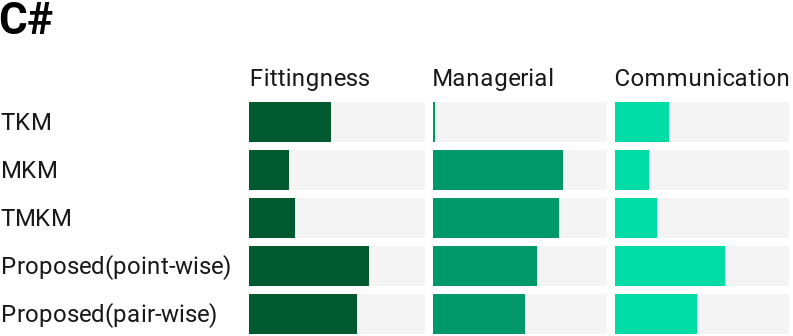}\hfill
\includegraphics[width=0.49\textwidth]{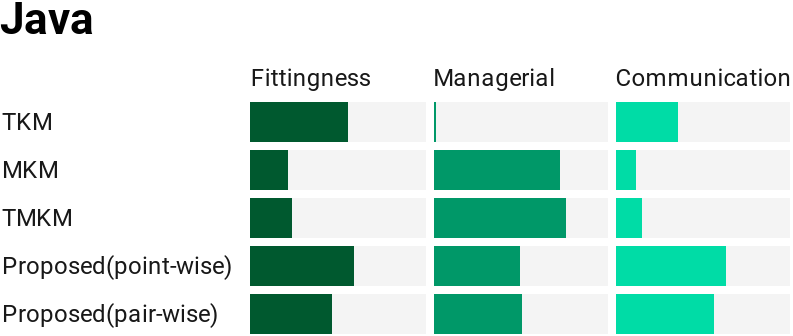}
\caption{NDCG@10 of all models with each leader relevancy metric as the grade, on C\# (left) and Java (right). Each bar spans a scale from 0 to 1.}
\label{fig:results}
\end{figure}

\subsubsection{Comparison of the Baselines (RQ1)}\label{sec:rq1}

The baseline rows of Figure~\ref{fig:results} show the NDCG@10 of the baselines with each leader relevancy metric as the grade. The two collections show the same pattern. TKM obtains the highest fittingness and communication scores among the baselines, since it only looks for intermediate knowledge in the project skill-areas. MKM and TMKM obtain the highest managerial scores, since both models include managerial knowledge, while TKM obtains almost none. Table~\ref{tab:main} reports NDCG@$k$ and GAP@$k$ with the F-measure as the grade. TMKM outperforms TKM and MKM on every metric in both collections, and the differences are statistically significant. Modeling technical and managerial knowledge together therefore balances the three qualities better than modeling either one alone.

\subsubsection{Comparison of the Negative Knowledge Models (RQ2)}\label{sec:rq2}

Table~\ref{tab:nkm} compares the NKM variants. Since NKM targets technical fit, we use fittingness as the grade in this table. K-NN obtains the best score on every metric in both collections. In addition, every point-wise model outperforms every pair-wise model on all four metrics in both collections. We use the best point-wise model (K-NN) and the best pair-wise model (SVM-Rank) in the full model, and we refer to the two configurations as \emph{Proposed (point-wise)} and \emph{Proposed (pair-wise)}.

\subsubsection{Comparison of the Proposed Model with the Baselines (RQ3)}\label{sec:rq3}

Table~\ref{tab:main} compares the proposed model with the baselines. Both configurations outperform all baselines on every metric in both collections, and the improvements are statistically significant. The point-wise configuration performs best. Compared with TMKM, it raises NDCG@10 from 0.236 to 0.621 on C\# and from 0.280 to 0.572 on Java. The proposed model extends TMKM with the negative knowledge penalty and the communication model, so these two components account for the gain. The point-wise configuration also outperforms the pair-wise configuration on every metric in both collections.

Figure~\ref{fig:results} breaks the comparison down by leader relevancy metric. Both configurations obtain higher communication scores than all baselines in both collections. The point-wise configuration also obtains the highest fittingness score in both collections, while the pair-wise configuration falls below TKM on Java. Both configurations give up part of the managerial score of MKM and TMKM. As Table~\ref{tab:main} shows, this trade-off still leads to the best balance of the three qualities.

\section{Discussion}\label{sec:discussion}

\subsection{Analysis of the Components}\label{sec:analysis}

Our model addresses two gaps of the baselines. The first gap is knowledge outside the project, which points to costly and over-involved candidates. NKM penalizes this knowledge, and the full model improves fittingness over the baselines in all cases except the pair-wise configuration on Java (Figure~\ref{fig:results}). Among the LTR methods, K-NN performs best. One explanation is that each feature corresponds to one skill-area of the negative query, so the distance between two feature vectors directly compares the knowledge of two candidates in each of these skill-areas.

The second gap is communication with the team. Following studies that place effective leaders at the center of their teams, LCM combines three centrality measures. The weights that CA learns (Section~\ref{sec:parameters}) are highest for degree centrality and next highest for closeness centrality in both domains. In our collections, the direct links between a candidate and the team members therefore carry most of the signal for communication. These weights may differ on other data.

\subsection{Comb-Shaped Candidates}\label{sec:comb}

\begin{figure}[t]
\centering
\includegraphics[width=0.9\textwidth]{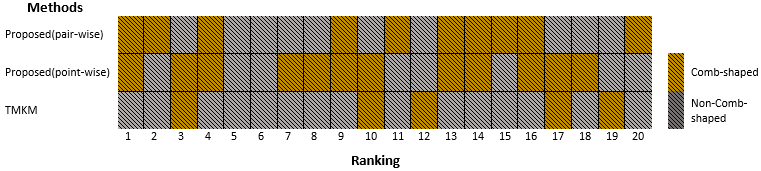}
\caption{Comb-shaped candidates among the top 20 candidates of the proposed model and of TMKM for a sample query in the Java domain.}
\label{fig:comb}
\end{figure}

In Section~\ref{sec:introduction}, we argued that the ideal leader is Comb-shaped, with intermediate knowledge in the technical skill-areas of the project and advanced managerial knowledge. Figure~\ref{fig:comb} shows the top 20 candidates of the proposed model and of TMKM, the strongest baseline, for one sample query in the Java domain. Gold cells mark Comb-shaped candidates who meet the leader conditions, and gray cells mark the other candidates. The point-wise and pair-wise configurations place 12 and 10 Comb-shaped candidates in the top 20, compared with 5 for TMKM.

\subsection{Practical Implications}\label{sec:practical}

Leader finding has direct use in hiring. Our experiments find leaders for software development projects and teams, where the choice of leader affects the outcome of the project \citep{shastri2021role,iriarte2020projects}. Our skill-area tree links skill-areas to team positions, which makes it straightforward to describe a project and a team as a query. The models only need documents that can be assigned to skill-areas and a way to measure relations between users. They therefore apply to other types of projects and data as well.

\subsection{Limitations}\label{sec:limitations}

Our study has several limitations. First, we derive knowledge levels from accepted answers, following prior work \citep{gharebagh2018t}, and these levels are an estimate of real expertise. Second, the communication metric uses shared skill-areas as an estimate of the ability to communicate, and we use the same metric to label the training data of CA. Cross-validation keeps the test queries out of training, but the learned weights still reflect this notion of communication. Third, our data covers Stack Overflow posts up to March 2015, and the managerial collection is much smaller than the technical collections. Fourth, a leader who meets our conditions does not guarantee the success of a project, since the team, the organization, time, and budget also matter. Finally, a leader has other qualities, such as responsibility and writing skills, that our model does not cover.

\section{Conclusion and Future Work}\label{sec:conclusion}

We studied leader finding, where the goal is to rank candidate leaders for a given project and team. We argued that a suitable leader fits the technical needs of the project without being overqualified, has advanced managerial knowledge, and communicates well with the team. Our probabilistic model scores technical fit with a negative knowledge penalty and communication with centrality measures on a communication graph, and we proposed one evaluation metric for each quality. On the Java and C\# collections of Stack Overflow, both configurations of the model outperform all baselines on every ranking metric.

In future work, we plan to replace the LTR models of NKM with deep learning models. Our method also has a high computational cost. We compute the graph and the shortest paths between all pairs of users offline, but every user of a collection outside the team is a candidate. Pruning candidates who are unlikely to lead the team before scoring would reduce this cost.

\bibliography{references}

\end{document}